# Information Transfer Time: The Role of Holomorphism, Stationary Phase, and Noise

**Michael C. Parker**

Fujitsu Network Communications Inc., Photonics Networking Laboratory, Richardson, Texas 75082, and Fujitsu Telecom Europe Ltd. Research, 38 Byron Avenue, Colchester, Essex, CO3 4HG, UK, Tel: +44 1206 542399, Fax: +44 1206 762916, Email: M.Parker@ftel.co.uk

**Stuart D. Walker**

University of Essex, Department of Electronic Systems Engineering, Wivenhoe Park, Colchester, Essex, CO4 3SQ, UK, Tel: +44 1206 872413, Fax: +44 1206 872900, Email: stuwal@essex.ac.uk

## Abstract

In this paper we present an analysis of information transfer time based on holomorphism, causality and the classical principle of stationary phase. We derive a common determination of transfer time for a range of pulse shapes and dispersive media in the absence of noise, and find our results are in accordance with those from the application of Laplace transform theory to networks. We also make a preliminary study of the effect of noise on information transfer time, and find that noise tends to increase transfer times. Noise and information signals are both essentially acausal, such that analytic continuation (i.e. prediction) is impossible, which also implies that their frequency spectra cannot be holomorphic. This leads to the paradox of a non-holomorphic information-bearing light signal, yet whose underlying Maxwell equations governing the propagation of the EM wave describe a holomorphic function in spacetime. We find that application of stationary phase and entropy arguments circumvents this difficulty, with stationary phase only suggesting the most likely transfer times of an information signal in the presence of noise. Faster transit times are not excluded, but are highly improbable. Stationary phase solutions, by definition, do not include signal forerunners, whose detection in the presence of noise is also unreliable. Hence a finite information capacity ensues, as expected from Shannon's law, and information cannot be transferred faster than $c$. We also find that the method of stationary phase implies complex transfer times. However, by considering spacetime to be isomorphic with the complex temporal plane, we find that an imaginary time is equivalent to a real distance, and can be interpreted as the uncertainty in the spatial position of the information pulse. Finally, we apply our theory to a photonic band gap crystal, and find that information transfer speed and tunneling is always subluminal.

## 1. Introduction

Recent reports of slow [1] and fast [2] light have excited interest in the photonics community due to the implications for optical information storage, optical delay lines, and the possibility of superluminal data transmission. In this paper, we shall examine these non-traditional forms of light from an information theoretic point of view, since this is how these phenomena may ultimately find useful application. Previous theoretical work (which has been extensively reviewed [3-5]) has tended to employ analytically-continuous Gaussian, or infinite-bandwidth step pulses to examine information velocity. However, Gaussian functions do not have a well-defined front, such that their speed of propagation becomes ambiguous. Also, infinite bandwidth signals cannot propagate through any real physical medium (whose transfer function is therefore finite) without pulse distortion, which also leads to ambiguities in determining propagation velocity.

The use of Gaussian pulses in a dispersive medium has been extensively studied [6-9], often with the assumption that all the information is in the leading edge of the pulse, thus allowing analytic continuation. Cauchy-Riemann theory has been extensively employed to analyse group velocity in dispersive media [10-12], with Chiao and co-workers also studying an inverted medium (amplifier), where adiabaticity requires that the leading edge is amplified, whilst the trailing edge is suppressed [13]. However, the authors of the report of pulse transmission at 310 times the speed of light employing Gaussian pulses [2], appear to disagree that superluminal velocity is simply due to amplification of the leading edge [14]. Diener [15] also makes a similar approach to Chiao, with an interesting discussion of Greens' function, and energy storage by the amplifying medium.

The use of a step-discontinuity to analyse signal speed was pioneered by Sommerfeld and Brillouin [10], who found that the step front travelled at the speed of light in vacuum $c$, via the forerunners, such that Einstein causality is not compromised. However, Nimtz [16] has pointed out that a step-discontinuity has infinite bandwidth, and so is not physically reasonable. Hence by inverting the argument, he makes the point that Einstein causality is only true for a well-defined front; but since this does not exist for a realistic time-limited signal, this might leave open the possibility of superluminal velocity [17]. From an information theoretic viewpoint, he also invokes Shannon's law [18], as a reason for a finite bandwidth requirement.

All analyses tend to yield complex time delays, ultimately due to a complex propagation constant. This has caused a great deal of confusion, with Landauer denying the physical reality to an imaginary time [19]. Analogous to the propagation of EM radiation through a dispersive medium is the tunnelling of an electron through a potential barrier [20]. After initial studies by Wigner [21], Büttiker has made important contributions to electron tunnelling times [22], and with Landauer, also to photon tunnelling times [23]. They make an interesting discussion of the compromise between a narrow bandwidth pulse (and therefore very long in spatial extent), and the position of the centre of gravity of pulse: an aspect of the uncertainty principle. Fertig also comes to similar conclusions [24], thus highlighting another problem with the "fast" light experiment [2]. Leavens and Aers [25,26] consider real and imaginary times to be simply "barrier interaction parameters with the dimensions of time", but do not give a physical meaning to complex time. They also discuss the additive properties of complex time. Sokolovski considers the concept of complex times to be an analogue to the Young's Twin Slit experiment [27]. Krausz and co-workers have performed experiments on photon tunnelling [28], and found that, in agreement with Hartman [29], tunnelling time does not depend on barrier thickness.

Recently, the effects of noise on signal velocity have also started to be examined [30], and Aharanov discusses quantum limitations and noise in superluminal propagation [31]. Finally, Chiao and co-workers have been using photon pairs to measure transfer times, but the stochastic nature of light becomes evident under these conditions, and questions of Einstein-Podolsky-Rosen (EPR) entanglement and the associated paradoxes make the results difficult to unambiguously interpret [32-36].

**2. Ubiquity of Holomorphic Functions**
In an important paper by Toll [37], it was found that a function, which is causal in time, has a holomorphic [38] Fourier transform (FT). That is to say, its frequency components, when analysed across the complex frequency plane, are holomorphic (i.e. analytic, or holographic) in nature, and obey the Cauchy-Riemann equations. Toll's paper can be further generalised, such that any function which is zero (or can be considered zero) from some point back to minus infinity on the real axis, will have a holomorphic FT. Thus any 'windowed' or 'apertured' function has a holomorphic FT. It is well known that the free-space far-field (Fraunhofer) diffraction pattern of light is mathematically equivalent to the FT of the original aperture. Hence, the resulting interference pattern from the Young's twin-slit experiment is holomorphic. Reflection and transmission coefficients are given by the FT of the finite spatial distribution of scatterers (bounded in spatial extent) [39], such that they too are holomorphic. Thus both transverse and longitudinal grating structures scatter light holomorphically.

The Maxwell equations are a form of the Cauchy-Riemann equations, such that EM radiation is holomorphic in spacetime. We demonstrate this for one space and one time dimension, although we begin with Maxwell's equations in 3+1 dimensions:

$$\nabla \cdot D = \rho \quad (1a) \qquad \nabla \wedge H = \frac{dD}{dt} + J \quad (1b) \qquad \nabla \wedge E = -\frac{dB}{dt} \quad (1c) \qquad \nabla \cdot B = 0 \quad (1d)$$

where the displacement is $D = \varepsilon E$, with $E$ being the electric field, and $\varepsilon$ is the permittivity; while $B = \mu H$ where $B$ is the magnetic flux density, $\mu$ is the permeability, and $H$ is the magnetic field strength; $\rho$ is the charge density, and $J$ is the current density. For an EM field in an isotropic dielectric medium, without any sources or free charges etc., we can assume that $\rho = J = 0$. Ignoring equations (1a) and (1d) which are therefore null, we can rewrite equations (1b) and (1c) in their 1+1-dimensional forms:

$$\frac{dH}{dx} = \frac{d\,\varepsilon E}{dt} \quad (2a) \qquad\qquad \frac{dE}{dx} = -\frac{d\,\mu H}{dt}. \quad (2b)$$

We rearrange equations (2) to yield:

$$\frac{d}{dx}\left(\sqrt{\frac{\mu}{\varepsilon}}H\right) = \sqrt{\varepsilon\mu}\frac{dE}{dt} \quad (3a) \qquad\qquad \frac{dE}{dx} = -\sqrt{\varepsilon\mu}\frac{d}{dt}\left(\sqrt{\frac{\mu}{\varepsilon}}H\right) \quad (3b)$$

If we define $c = 1/\sqrt{\varepsilon\mu}$ (i.e. speed of light), and $Z = \sqrt{\mu/\varepsilon}$ (i.e. impedance of the medium), and for convenience $A = HZ$, and $y = ct$, then we can straightforwardly rewrite equations (3) as:

$$\frac{dA}{dx} = \frac{dE}{dy} \quad (4a) \qquad\qquad \frac{dE}{dx} = -\frac{dA}{dy}, \quad (4b)$$

which are the classical 2-dimensional form of the Cauchy-Riemann equations. Hence we can assume that the electric field $E$ and the magnetic field $H$ (which is simply proportional to $A$) together form a holomorphic function[1], so that the overall holomorphic function is $F = E + jHZ$, and the space and time coordinates make up the complex plane $z$, such that $z = x + jct$.

In our physical universe, negative frequencies cannot be directly measured. Only by frequency-shifting (i.e. multiplexing onto a positive high-frequency carrier wave) do they become positive, so that they can be (somewhat indirectly) 'observed'. We have here an analogous 'frequency-causality' in the frequency domain, where the frequency spectrum is immeasureable (to all intents equal to zero) for negative frequencies. Since the EM signal is holomorphic in spacetime due to the Maxwell equations, Toll's paper [37] requires it to be 'causal' in its (inverse) FT space, i.e. in its frequency (reciprocal, or Hilbert) space. Hence, we can see that the holomophic Maxwell equations, by definition, require the non-measurability of negative frequencies. In passing, we note that the Hilbert space of 2-D spacetime is defined by a spatial-frequency on the real axis, and temporal-frequency along the imaginary axis. Thus we now notice that a causal EM wave is both holomorphic in the complex frequency plane, and also in spacetime. Applying Occam's Razor to the situation in a quest for simplicity, we make the complex frequency plane, and the Hilbert space of 2-D spacetime isomorphic with one another; and hence also make spacetime equivalent to the complex time plane. In which case, without altering the underlying mathematical arguments above, we find that an imaginary time is equivalent to real space, with a similar equivalence for their reciprocal spaces.

Finally, we note that holomorphic functions contain within themselves the essence of a wave-particle duality, since they have both global and local properties. Holomorphic functions are completely defined over the whole complex plane; yet can also be fully defined from any arbitrary point in the complex plane, by use of the Cauchy integral theorem and summation of an unconditionally convergent Taylor's series [40]. Thus any single point in the complex plane contains all the information required to completely reconstruct the holomorphic function over all space. Dual (i.e. complementary) descriptions to fully characterise a holomorphic function thus exist, and the similarities to a "hologram" are obvious. Hence we see that the Maxwell equations, although developed as a set of EM wave equations, because of their holomorphic nature, also implicitly describe a localised photon in spacetime.

**3. Non-Holomorphic Functions: Information and Noise**
A holomorphic function has the property that it is infinitely redundant, i.e. the function can be defined over all space merely by appropriate continuation (i.e. by infinite differentiation, and infinite Taylor's series) of an arbitrary point on the function. A Gaussian function is holomorphic, which is why its 'leading edge' can be used to define the complete function; hence the explanation for the "superluminal" velocity observed for such a signal in an inverted medium: the principle of analytic continuity. Thus we argue that no information transfer actually takes place when a holomorphic signal is used to 'transmit' information from A to B, since such a signal is completely defined over all space and so is already completely defined at the destination point B. Hence there is a zero time delay (and infinite velocity) to 'transfer' the signal. Instead, only non-holomorphic signals can be used to transmit information, since they are by definition not defined over all space. This is intuitively reasonable, since we often associate an information signal with a discontinuous function, which cannot be analytic and allow 'prediction'.

By inverting the argument of Toll's paper [37], a non-holomorphic information-bearing signal contains the implicit assumption of acausality. However, noise also has this characteristic, since it is unpredictable, and tends to be acausal (e.g. spontaneous emission); but it contains no information. So, how do we tell the difference between pure noise, and a high-information bearing signal, since both signals have the superficial appearance of noise? As can be derived from Gödel's incompleteness theorem, and Turing's halting/indecidability problem [41], we find that there can exist no finite and deterministic algorithm, which can decide whether any given signal is random (i.e. purely noise), or rich in information. In addition, the amount of information in a given signal cannot be deterministically calculated [42,43], which further makes a definitive information transfer rate or time difficult to assess.

---

[1] As might be expected, the full 3+1-dimensional Maxwell equations also form the appropriate Cauchy-Riemann equations for a 4-dimensional manifold.

Since the solutions to wave propagation equations tend to involve FT quantities, we find that wave function solutions are holomorphic in nature. A wave equation solution can also be regarded as the distribution of amplitudes and phases (i.e. the probability distribution) indicating the range of state possibilities available to a particle. However, when a measurement is made on the position (for example) of an individual particle, *after* the measurement, the particle is found in only one of those possible states. The 'distribution' function of that particle after the measurement is now similar to a Dirac delta function – it has been localised. The wave function has collapsed, or decohered after the measurement [44]; and as a result it is also now non-holomorphic. The decohered, non-holomorphic wave function now bears information, in the form of telling us where the particle is situated [45]. Thus a measurement is associated with the transformation of a holomorphic function into a non-holomorphic function. In so doing, information is exchanged.

Thus information has analogies with a field or particle, since it can propagate, and be exchanged. Information also tends to be associated with entropy [46-48]. They are often considered to be negatively correlated, e.g. Brillouin's concept of negentropy (i.e. negative entropy) as an alternative measure for information [49]. We can therefore ascribe a physical difference between 'real' and 'virtual' processes: A real process (such as a measurement) is associated with an exchange of information and a change in entropy; whereas a virtual process is associated with no information exchange and sees no change in entropy. In many of the 'superluminal' experiments employing Gaussian pulses, adiabaticity is assumed due to the smoothness of the pulse, such that there is no change in entropy; in which case no information is transferred. As noted above, an information-bearing signal will tend to have discontinuities or abrupt changes, such that non-adiabatic processes occur, and entropy changes accompany the pulse through an active medium. This is in accord with the concept that reversible computation, i.e. processing (without loss) of information, doesn't require energy or a change in entropy, whereas erasure of information requires energy and is accompanied by a change in entropy [48]. Transfer of information from A to B can be understood to be the annihilation of the information at A, proceeded by the re-creation of that same information at B.

**4. Stationary Phase and Holomorphic Functions**
The principle of stationary phase to describe the group velocity is well known, and was employed by Sommerfeld and Brillouin in their early studies of wave propagation and group velocity [10]. The principle of stationary phase can also be used in a statistical sense, such that the most likely events tend to be associated with slowly-varying phase variation in the frequency domain, and unlikely events tend to be associated with rapidly varying phase with frequency. Hence the principle of stationary phase can be used to predict the most likely outcome of an experiment, but it doesn't exclude the possibility of alternative outcomes. When applied to information transfer, it predicts the most likely speed at which information can be transmitted, but allows the possibility of superluminal information velocity, albeit with only a low probability. Trying to transmit information at such superluminal speeds therefore necessarily means a severe degradation in the quality of that information, since only a small proportion of that information will be successfully transmitted. In which case, error-correcting techniques need to be employed to reconstruct the information at the far end. This is at the expense of the apriori requirement of a large degree of redundancy in the original information signal (i.e. akin to making it holomorphic!), and also processing time at the receiving end; in which case it is questionable whether truly superluminal information transmission has then been achieved.

When considering the principle of stationary phase along the real frequency axis, we assume that a wave can be written as $\exp j(\omega t - \beta z)$, such that the phase of the wave is given by $\varphi = \omega t - \beta z$, and the propagation constant is given by $\beta(\omega) = n(\omega)\omega/c$, where $n(\omega)$ is the frequency-dependent (and holomorphic [38]) refractive index, and $c$ is the wave velocity in vacuum (i.e. speed of light in vacuo.) When the phase is stationary with respect to frequency, such that $\partial\varphi/\partial\omega = 0$, the group velocity is given by:

$$v_g = \frac{z}{t} = \frac{\partial\omega}{\partial\beta}, \qquad (5)$$

and the time taken to propagate a distance $z$ through the medium is given by:

$$\tau_n = z\frac{\partial\beta}{\partial\omega} = \frac{z}{c}\left(n + \frac{\partial n}{\partial\omega}\omega\right). \qquad (6)$$

After propagating a distance $z$ in the medium, the phase accumulated by the wave is $\phi_n = -\beta z$. Compared with propagation in vacuum, which would accumulate a phase $\phi_0 = -\omega z/c$ and take a time $\tau_0 = z/c$, there is a relative time delay $\Delta\tau$ given by:

$$\Delta\tau = \tau_n - \tau_0 = z\left(\frac{\partial\beta}{\partial\omega} - \frac{1}{c}\right) = \frac{z}{c}\left(\frac{\partial n}{\partial\omega}\omega + n - 1\right). \qquad (7)$$

When $\Delta\tau < 0$, the wave is travelling at a superluminal velocity (i.e. "fast" light), and we have "slow" light when $\Delta\tau > 0$. We can also describe a relative phase $\Delta\phi = \phi_n - \phi_0$ whose derivative with respect to frequency is the relative time delay [12]:

$$\Delta\phi = -\frac{z}{c}\left(n(\omega) - 1\right)\omega \qquad (8) \qquad \text{such that} \qquad \Delta\tau = -\frac{\partial\Delta\phi}{\partial\omega}. \qquad (9)$$

### 4.1 Lossy & Amplifying Media, and Classically Forbidden Regions

For a complex-valued propagation constant $\beta = \beta_r + j\beta_i$, associated with a complex-valued refractive index (e.g. for lossy and amplifying media, or a classically forbidden region), equation (5) cannot be directly evaluated (except via a Jacobean which only gives a modulus result), and equation (7) yields a complex-valued relative time delay. Sommerfeld and Brillouin used contour integration in the complex frequency plane to analyse the propagation of a signal through an absorptive medium. The complex frequency plane is described by $\acute{\omega} = \omega + j\upsilon$, where $\omega$ is the (conventional) frequency along the real axis, and $\upsilon$ is frequency along the imaginary frequency axis. The phase of the propagating wave is thus a function of $\acute{\omega}$ in the complex frequency plane, and consists of both real and imaginary components:

$$\varphi(\omega,\upsilon) = \varphi_r(\omega,\upsilon) + j\varphi_i(\omega,\upsilon), \qquad (10)$$

where $\varphi_r$ and $\varphi_i$ are purely real functions. Due to considerations of causality, $\varphi_r$ and $\varphi_i$ are Hilbert transforms of each other, such that $\varphi(\omega,\upsilon)$ is itself a holomorphic function. For a purely transparent medium (e.g. for vacuum) the imaginary component is zero, and the conventional stationary phase approach degenerates to $\partial\varphi/\partial\omega = \partial\varphi_r/\partial\omega = 0$. However, applying the principle of stationary phase to the real phase component, with respect to the two orthogonal frequency axes (i.e. method of saddle points [10]) requires:

$$\frac{\partial\varphi_r}{\partial\omega} = 0 \qquad (11a) \qquad \text{and} \qquad \frac{\partial\varphi_r}{\partial\upsilon} = 0. \qquad (11b)$$

These two differentials are in turn associated with two relative group delays, a real one, and an imaginary one in the complex time plane:

$$\Delta\tau_r = -\frac{\partial\Delta\phi_r}{\partial\omega} \qquad (12a) \qquad \text{and} \qquad \Delta\tau_i = -\frac{\partial\Delta\phi_r}{\partial\upsilon}, \qquad (12b)$$

where equation (12a) is equivalent to (9) for a purely transparent medium. In the same way that $\varphi(\omega,\upsilon)$ is an analytic function, the relative phase response function $\Delta\phi(\omega,\upsilon) = \Delta\phi_r + j\Delta\phi_i$ is also holomorphic in the complex frequency plane $\acute{\omega}$. The overall relative time delay is then given by:

$$\Delta\tau = \Delta\tau_r + j\Delta\tau_i. \qquad (13)$$

### 4.2 Holomorphic Transfer Function of a Causal Medium

For a causal medium, the transmission frequency response $t(\omega) = |t(\omega)|e^{j\Delta\phi_r(\omega)}$ is a holomorphic function, where the real relative phase response $\Delta\phi_r = \phi_n - \phi_0$ is the phase compared with that of a wave travelling the same geometric distance in vacuum $\phi_0$. Taking the natural logarithm of $t(\omega)$ also yields a holomorphic function, such that $\ln|t(\omega)|$ and $\Delta\phi_r(\omega)$ are Hilbert transforms of each other, and obey the Cauchy-Riemann equations in the complex frequency plane $\acute{\omega}$:

$$\frac{\partial}{\partial\omega}\Delta\phi_r(\acute{\omega}) = \frac{\partial}{\partial\upsilon}\ln|t(\acute{\omega})| \quad (14a) \qquad \text{and} \qquad \frac{\partial}{\partial\upsilon}\Delta\phi_r(\acute{\omega}) = -\frac{\partial}{\partial\omega}\ln|t(\acute{\omega})|. \qquad (14b)$$

We notice that the left-hand sides of the Cauchy-Riemann equations (14a) and (14b) are also equivalent to the relative time delay functions (12a) and (12b) respectively. In which case, comparing equation (14b) and (12b), we must have that the imaginary relative group delay is also given by:

$$\Delta\tau_i = -\frac{\partial\Delta\phi_r}{\partial\upsilon} = \frac{\partial}{\partial\omega}\ln|t(\acute{\omega})|, \quad (15a) \qquad \text{and therefore also} \qquad \Delta\tau_r = -\frac{\partial\Delta\phi_r}{\partial\omega} = -\frac{\partial}{\partial\upsilon}\ln|t(\acute{\omega})|. \qquad (15b)$$

These dual relationships for each of the real and imaginary time delays have already been observed by Leavens and Aers [26], but they did not comment on the fact that they are simply an example of the Cauchy-Riemann equations. Hence substituting (12a) and (15a) into (13), the overall complex relative group delay is given by:

$$\Delta\tau = -\frac{\partial\Delta\phi_r}{\partial\omega} + j\frac{\partial}{\partial\omega}\ln|t(\omega)|. \tag{16}$$

We can straightforwardly rewrite equation (16) as:

$$\Delta\tau = j\frac{\partial}{\partial\omega}\ln[t(\omega)]. \tag{17}$$

Thus the relative time delay is simply given by the frequency-derivative of the logarithmic complex transfer function of the medium. We have derived this expression for the relative time delay using only the principle of stationary phase along both real and imaginary frequency axes, and invoking causality. Often, only the delay due to the variation of phase with frequency is calculated for the group delay, as the transmission amplitude at the frequency of interest is often almost constant, or is at a maximum (i.e. as close to 100% as possible), e.g. the transmission of a Fabry-Perot resonator, or a low-loss data transmission line. Conventional complex-number calculations of phase variation with frequency (and hence time delay) make the implicit assumption of monochromatic stimulation, which is non-information bearing. However, analysis of the dispersive character of any real medium shows straightforwardly that non-monochromatic signals suffer varying time delay. This translates directly to pulse shape distortion in the time domain (the essence of dispersion). Chirped-frequency evaluation of media is also well known to produce distorted envelopes as a function of chirp time. These common-place observations support the arguments advanced for an imaginary time delay component, based on amplitude variation as well as phase variation.

The complex time delay produced using equation (17) has the conventional time delay as its real part, but also an interesting Hilbert transform pair as the imaginary part. Analogous equations have been derived for electron tunnelling times associated with a low probability of transmission, such that Leavens and Aers [26] have also written the transfer time as:

$$\Delta\tau = j\hbar\frac{\partial}{\partial E}\ln(|t|e^{j\Delta\phi}), \tag{18}$$

where $E = \hbar\omega$ is the electron energy. Büttiker has described the tunnelling (transmission) time in modulus format [22]:

$$\Delta\tau = \frac{m}{\hbar\kappa}\left|\frac{\partial}{\partial\kappa}\ln(|t|e^{j\Delta\phi})\right|, \tag{19}$$

where $m$ is the electron mass, $\kappa^2 = k_0^2 - k^2$ with $k_0 = \sqrt{2mV_0}/\hbar$ being the wave-number of potential barrier $V_0$, $k$ being the particle wave-number, and $t = |t(\kappa)|e^{j\Delta\phi(\kappa)}$ is the complex forward scattering probability (or transfer function) of the barrier. Likewise, the time delay for photon tunnelling has also been given as [23]:

$$\Delta\tau = \frac{\partial\kappa}{\partial\omega}\left|\frac{\partial}{\partial\kappa}\ln(|t|e^{j\Delta\phi})\right|. \tag{20}$$

Equation (18) is in broad agreement with (17)[2], however equations (19) and (20) are each expressed in modulus format, which circumvents the complex-time aspect.

### 4.3 Dependence of Time Delay on Noise
From filter theory the time (group) delay is given by $\Delta\tau = -\partial\phi/\partial\omega$, but this is generally under the implicit assumption of a lossless and noiseless channel. It might be expected that a lossy and noisy channel will exhibit a different group delay characteristic. There are various aspects to defining the time taken to transmit a signal through a channel. Using heuristic arguments, we demonstrate that the definition of time taken to transmit a signal (consisting of a single pulse) a distance $L$ from A to B depends on the amount of noise in the channel, assuming finite channel transmission:

 a) If there were no noise, then theoretically speaking, a detector could unambiguously detect a signal by the presence of the Sommerfeld/Brillouin forerunners, which travel at a speed $c$ through the channel. Thus in this case, the time delay to transfer information would simply be $\Delta\tau = L/c$, since we can assume that $t(\omega) = \exp(j\omega L/c)$.
 b) If there were some system noise, then the earliest time for 'unambiguous' detection of a pulse of information would be when the amplitude of the signal first rises above that of the noise. This is a slower time compared with (a), and would be closer to, but still less than $\Delta\tau \leq nL/c$, where $n$ is the refractive index of the medium.
  c) If there were moderate noise (the usual case), then the detector would have to wait until the centre of mass of the pulse had arrived, in order to unambiguously detect the pulse, in which case the 'stationary phase' condition applies, and the time delay is $\Delta\tau = -L[\partial\omega/\partial\beta_r]^{-1} = nL/c$.
  d) If the noise power were much greater than the signal power, then the detector/decoder has increasing difficulty in being able to unambiguously deliver correct information, and the time delay is $\Delta\tau \to \infty$.

---
[2] We note that in the original paper [26] there is a sign change in front of the first '$j$' term in equation (18), due to their convention of defining the phase as $\varphi = \beta z - \omega t$.

We note that there are also timing information issues, which we have ignored, were the signal to consist of a sequence of pulses. If the signal power is reduced such that the signal-to-noise ratio (SNR) is reduced, then the time delay will also tend to increase, so that we expect an attenuating channel to delay a signal by an extra amount. Thus the amplitude response of the channel (as well as its phase response) determines the time delay of a signal, as indicated by equation (17). A slower information transfer time due to the presence of noise is equivalent to a reduction in the channel transmission bandwidth, since the quantity of useful information transmitted per unit time is therefore lowered. This is in agreement with Shannon's information capacity theorem [18].

**5. Application of Theory**
**5.1 Simple Filter Transmission Functions**
To indicate the utility of equation (17), we apply it to some well-understood generalised low-order frequency response functions. Unfortunately no propagation constant or length parameter is associated with these transmission functions as they are highly abstracted. This makes it is difficult to assign an appropriate real length with the imaginary time delay for these network functions, since they do not have any physical size parameters. However, in general, the temporal distribution of a pulse can be straightforwardly regarded as a physical length by involving the propagation velocity of the pulse. In which case, the imaginary time delay can be understood to indicate uncertainty in the spatial pulse position through the network. In the next section 5.2, we consider light transmission through a photonic crystal, which does have complete physical parameterisation. A first order low-pass filter can be simply described by the function:

$$t(\omega) = \frac{1}{1 + j\omega\tau_1}, \tag{21a}$$

where $\omega = 1/\tau_1$ is the cut-off frequency associated with the filter. Applying equation (17) to (21a) allows us to express the time delay as:

$$\Delta\tau(\omega) = \frac{\tau_1}{1+\omega^2\tau_1^2} - j\frac{\omega\tau_1^2}{1+\omega^2\tau_1^2} \tag{21b}$$

For low frequencies, well below cut-off, the time delay is close to $\tau_1$ (as expected from Laplace transform theory) and the imaginary component is negligible. However, at the cut-off frequency associated with the filter, $\omega = 1/\tau_1$, the time delay is given by:

$$\Delta\tau_{\omega=1/\tau_1} = \frac{\tau_1}{2}(1-j), \tag{21c}$$

such that the imaginary time delay is as large as the conventional time delay. In fact, the imaginary time delay is maximised at the cut-off frequency (i.e. the band edge), which coincides with a region of high dispersion. In general, regions of high dispersion are associated with large imaginary time delays. As the frequency increases towards infinity, and the transmission becomes ever closer to zero, the imaginary component to the group delay dominates over the real time delay, yet also decreases inversely proportionally to the frequency. At these high frequencies, there is very little probability of transmission, but the time delay associated with any transmitted portion of the wave is also extremely small, such that the propagation is almost as if it where through vacuum. However, the real time is always positive for all frequencies, in which case the transfer time is always subluminal.

A $2^{nd}$-order bandstop filtering action with a finite $Q$ value (analogous to that of a finite photonic crystal described below in section 5.2) is described by

$$t(\omega) = \frac{1-\omega^2\tau_2^2 + j\omega\tau_3}{1-\omega^2\tau_2^2 + j\omega\tau_1} \tag{22a}$$

where $\tau_1$ is associated with the damping coefficient (or loss coefficient of the medium) and defines the width of bandstop, $\omega = 1/\tau_2$ is frequency at the centre of the stopband, and $\tau_3$ defines the attenuation of the bandstop. Applying equation (17) to (22a) allows us to express the time delay as:

$$\Delta\tau(\omega) = \frac{\tau_1(1+\omega^2\tau_2^2)}{(1-\omega^2\tau_2^2)^2+\omega^2\tau_1^2} - \frac{\tau_3(1+\omega^2\tau_2^2)}{(1+\omega^2\tau_2^2)^2+\omega^2\tau_3^2} - j\omega\left[\frac{\tau_1^2 - 2\tau_2^2(1-\omega^2\tau_2^2)}{(1-\omega^2\tau_2^2)^2+\omega^2\tau_1^2} - \frac{\tau_3^2 - 2\tau_2^2(1-\omega^2\tau_2^2)}{(1-\omega^2\tau_2^2)^2+\omega^2\tau_3^2}\right]. \quad (22b)$$

At low frequencies close to d.c., the time delay is purely real and given by $\tau_1 - \tau_3$ (in agreement with Laplace transform theory). If $\tau_3$ had a larger magnitude than $\tau_1$, then superluminal speeds (i.e. greater than $c$) would be implied at low frequencies. However, $\tau_3$ is always less than $\tau_1$ for any physically realisable medium. At the frequency associated with the bandstop centre of the filter, $\omega = 1/\tau_2$, the imaginary component to the time delay is zero, while the real part is given by $\Re\Delta\tau_{\omega=1/\tau_2} = 2\tau_2^2(1/\tau_1 - 1/\tau_3)$. This real part is found to be negative in the bandstop region, which is well known. However, it is only negative due to the lack of a propagation constant associated with the transfer function (22a). If a propagation constant equivalent to $\beta \equiv 2\tau_2^2(\tau_1-\tau_3)\omega/(\tau_1\tau_3 L)$ is included in equation (22a), with a putative length $L$ assumed for the network, then the overall real group delay is always greater than or equal to zero. As the frequency gets large and tends towards infinity, the real component of the time delay of (22b) tends towards zero as $\omega^{-2}$, whilst the imaginary component tends towards zero with only $\omega^{-1}$, such that the propagation is almost as if it where through vacuum. However, as for the first-order filter, the real time remains positive for all frequencies, such that the transmission time is always subluminal.

Transfer functions (21a) and (22a) can be straightforwardly rewritten in the Laplace $s$-plane, where $s = j\omega$, and the inverse Laplace transform determined, which essentially gives the causal time domain response. However, the resulting temporal pulse shapes demonstrate an ambiguity in determining the appropriate time delay, since depending on the relative damping of the medium, the leading edge or precursor can show non-monotonic behaviour. Application of equation (17) to the transfer function in the complex-frequency plane yields an unambiguous time delay for each frequency. A pulse with a temporal distribution $p(t)$ has a frequency spread $P(\omega)$, so that we can define a descriptor for the spectrally-averaged group-delay as:

$$\overline{\Delta\tau} = \frac{\int_{-\infty}^{\infty} \Delta\tau(\omega)|P(\omega)|^2 d\omega}{\int_{-\infty}^{\infty} |P(\omega)|^2 d\omega}. \quad (23)$$

For a pulse consisting of low-frequency components close to d.c., which is transmitted through a medium characterised by the low-pass filter of equation (21a), we can see that the average group-delay is close to $\overline{\Delta\tau} \approx \tau_1$. However, if the pulse consists of only high-frequency components, well above $1/\tau_1$, the average group-delay is close to zero, such that the pulse velocity is close to $c$, but the transmission is also close to zero. In which case, the pulse can be considered to be tunnelling with very low probability through the medium. The average real group delay is still always positive, such that the tunnelling still occurs at velocities less than $c$.

In passing, we note that the holomorphic nature of a causal transmission function $t(\omega)$ implies that the amplitude and phase responses must always be varying functions of frequency. This is in an analogous fashion to the property of light, which by definition cannot be halted in spacetime, according to the holomorphic Maxwell equations (and also Einstein's axiom.) Hence the concept of an all-pass filter (APF), which assumes 100% (constant) transmission over a continuous band of frequencies, is physically unrealisable.

**5.2 Photonic Crystals**

Figure 1 shows a 1-D photonic crystal (PC), with a photonic band gap centred on frequency $\omega_B$, corresponding to a Bragg wavelength $\lambda_B = 2\pi c/\omega_B$. We assume a constant reflection coupling-coefficient $\kappa_B = (\varepsilon_1 - \varepsilon_2)/\overline{n}\lambda_B$ along the finite length $L$ of the PC, with relative permittivity contrast $(\varepsilon_1 - \varepsilon_2)$, and average refractive index $\overline{n} = \sqrt{(\varepsilon_1+\varepsilon_2)/2}$, such that the equation describing the transmission through the structure is [51]:

$$t = \text{sech}[\kappa_B L \,\text{sinc}(sL)]\exp[-j\beta L], \quad (24)$$

where the propagation constant $\beta$ of the wave within the photonic crystal is given by:

$$\beta = \frac{\pi}{\Lambda} + \delta - \tilde{\kappa}(\delta) \sqrt{1 - e^{-\left(\frac{\delta}{|\tilde{\kappa}(\delta)|}\right)^2}} \pm j\tilde{\kappa}(\delta) e^{-\frac{1}{2}\left(\frac{\delta}{|\tilde{\kappa}(\delta)|}\right)^2} \quad (25)$$

As would be expected, the propagation parameter of the photon wave within the photonic crystal has both real and imaginary components, with the imaginary component maximised at the Bragg resonance, when the wave is at its most evanescent. From symmetry considerations, the $\pm$ symbol in equation (25) is related to the phase conjugate terms which are required to form the upper and lower bands in the Brillouin zone. The conventional detuning parameter, related to the PC period $\Lambda$ is given by

$$\delta = \bar{n}\omega/c - \pi/\Lambda, \quad (26)$$

such that $\delta = 0$ at the Bragg resonance. Taking the high dielectric contrast in account, the modified detuning parameter $s$, is given by:

$$s^2 = \delta^2 - |\tilde{\kappa}|^2 (1 - \exp[-(\delta/|\tilde{\kappa}|)^2]). \quad (27)$$

The coupling coefficient in the frequency domain is given by:

$$\tilde{\kappa}(\delta) = \kappa_B \mathrm{sinc}(\delta L) e^{-j\delta\Lambda}, \quad (28)$$

such that it is equal to $\kappa_B$ at the Bragg resonance. Applying equation (17) to equation (24), and ignoring any frequency dependency of the refractive index, the resulting real and imaginary group delays for the photonic crystal are plotted in figure 2. The associated transmission function (24) is also shown in figure 2. We have chosen a PC operating in the near-infrared regime at a favoured telecommunications wavelength of 1.5μm, with a Bragg resonance at $f_B = 200\mathrm{THz}$, with refractive indices $n_1 = \sqrt{\varepsilon_1} = 2.5$ and $n_2 = \sqrt{\varepsilon_2} = 1.5$, and period $\Lambda = \lambda_B/2\bar{n} = 364\mathrm{nm}$, and $M=5$ periods, such that the grating strength is given by $\kappa_B L = 2.35$, where the PC length is given by $L = M\Lambda$.

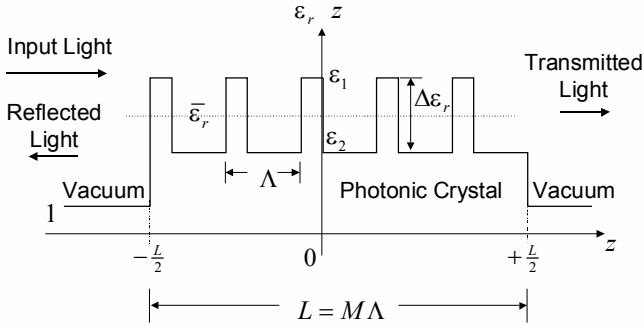

Figure 1: High dielectric contrast 1-D regular PC grating.

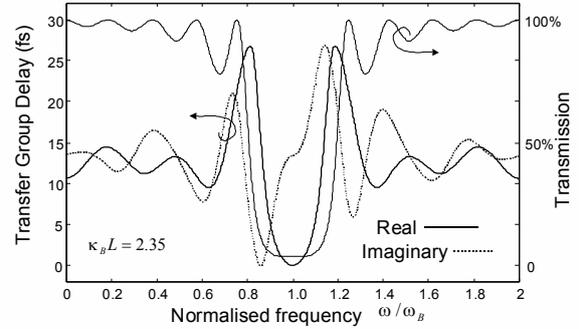

Figure 2: Real and imaginary group delays for 5-period photonic crystal.

By considering the real part of the propagation constant in equation (25), it is straightforward to demonstrate that the real group delay close to the Bragg resonant frequency is given by:

$$\Delta\tau_B = L\frac{\partial\beta_R}{\partial\omega} = \frac{\bar{n}L}{c}\left(1 - \frac{|\delta/\kappa_B| e^{-(\delta/\kappa_B)^2}}{\sqrt{1 - e^{-(\delta/\kappa_B)^2}}}\right). \quad (29)$$

Of interest is that equation (29) indicates that for any finite coupling strength $\kappa_B$, at the Bragg resonance when $\delta = 0$, the real group delay is always $\Delta\tau_B = 0$, i.e. the group delay at resonance is always that of vacuum propagation. However, we also note that off-resonance the group velocity is always less than $c$, since the group delay in equation (29) is always positive for detuning from resonance. This is also seen in figure (2). The imaginary group delay is zero at the Bragg resonance, but for clarity has been shifted to the mid-way value of the real group delay in figure 2. Thus we can see that its departures from zero are very close to the same magnitude as the departures of the real group delay from its mid-value. In this particular example, the imaginary group delay is always less than $|\Delta\tau_i| \leq 13.5\mathrm{fs}$. Given that the length of the PC is $L = 1.82\,\mu\mathrm{m}$, and the average refractive index is

$\bar{n} = 2.06$, we would expect light to require a non-resonating propagation time of 12.5fs. By associating the imaginary time delay with real space via $c/\bar{n}$, we have that 13.5fs is associated with a length of $L = 1.97$ μm, i.e. close to the length of the PC. When the light is highly dispersed by the PC, then the imaginary time delay indicates the degree of uncertainty in spatial position within the PC as the pulse is smeared out. As the coupling strength becomes very weak, equation (29) for the group delay becomes:

$$\tau_B = \frac{\bar{n}L}{c}[1 - \delta_{Dirac}(\delta)] \qquad (30)$$

where $\delta_{Dirac}(\delta)$ is the Dirac delta function which is zero when $\delta \neq 0$, and has unity weight for $\delta = 0$. An information-bearing signal has a finite bandwidth, and since the group delay is zero for only the Bragg resonant frequency, but increases either side of the centre frequency, it follows that information can only propagate at an average speed less than *c*. For a very weak band-gap, then the average group delay for an information-bearing signal (i.e. the frequency average of equation 30) is as expected $\bar{n}L/c$. Although we have performed a noiseless (best-case) analysis, we can assume that the presence of noise will slow the signal still further. Hence, for a photonic band-gap, where tunnelling occurs, we find that information transfer speed is always subluminal.

**Conclusions**

In this paper we have considered information transfer times over perspectives ranging from holomorphic functions, filter transfer functions, and Shannon capacity theory, to standard causal Laplacian formalisms. A stationary phase approach has been shown to be beneficial in resolving the paradox of information bearing and holomorphic functions in the presence of noise. We also conclude that information transfer is accompanied by a change in entropy of the system. In the general case, dispersive effects in a causal medium manifest themselves by the presence of an imaginary component to the group delay, which conspires to prevent accurate measurement of transit time. The real and imaginary time delays make a Hilbert transform pair, such that the imaginary time delay can be physically understood to be a real spatial length in spacetime. In relatively non-dispersive regions, such as the extrema and band-centre of resonant media, we have found that a positive real time delay is always predicted. We have studied light propagation through a photonic band gap device, and have shown that superluminal velocity (i.e. faster than conventional speed of light in the medium, *c/n*) is indeed possible in the resonant condition, but with two caveats: firstly the transmission (or tunnelling probability) is negligible; and secondly the maximum velocity can still only approach that of vacuum, but not exceed it. We also note that the presence of noise in the transmission system will tend to slow the information transfer time, in accordance with Shannon's theorem.